# Potential reduction interior point algorithm unfolding of neutron energy spectrum measured by recoil proton method

WANG Guanying [a], HAN Ran [a, b*], OUYANG Xiaoping [a, c], HE Jincheng [a], YAN Junyao [a]

[a] North China Electric Power University, Beijing 102206, China

[b] Beijing Institute of Spacecraft Environment Engineering, Beijing 100029, China

[c] Northwest Institute of Nuclear Technology, Xi′an Shaanxi 710024, China

## Abstract

A reconstruction algorithm for unfolding the neutron energy spectrum has been developed based on the potential reduction interior point. This algorithm can be easily applied to the neutron energy spectrum reconstruction in the recoil proton method. We transform neutron energy spectrum unfolding problem into a typical nonnegative linear complementarity problem. The recoil proton energy spectrum and response matrix at angles of 0° and 30° are generated by the Geant4 simulation toolkit. Several different neutron energy test spectrums are also employed. It is found that this unfolding algorithm is stable and provide efficiency accurate results.

---

[*] Corresponding author. E-mail address: hanran@ncepu.edu.cn

## 1. Introduction

Neutron energy spectrum is of great significance in both experimental research and theoretical analysis. It carries a large amount of information about the characteristics of the nuclear reaction system [1]. Recoil proton detectors (e.g. those based on the liquid scintillators and Si(Au) surface barrier detectors) are often used for the measurement neutron energy spectrum [2-4]. Recoil proton method is a common method for the diagnosis of neutron energy spectrum. For the occasions of low intensity and wide pulse width radiation field of high resolution measurement which other methods have difficult [5], it is widely used.

Neutrons incident polyethylene target after collimation, elastic scattering of hydrogen nuclei in polyethylene film and neutron [6]. Some of recoil proton enter into the detector which set in a certain angle. The direct recording of the detector is the recoil proton energy spectrum. The process of unfolding is to obtain the real neutron energy spectrum through the recoil proton energy spectrum and the response function.

To unfold the neutron energy spectrum, several mathematical methods and computing codes have been developed, such as maximum entropy deconvolution [7], genetic algorithm [8], GRAVEL code [14], and populated artificial neural networks(ANNs) [9-10]. GRAVEL is developed by PTB Laboratory. It is widely used in the field of energy spectrum unfolding. Maximum entropy deconvolution, genetic algorithm, and ANNs are the newly developed algorithm in recent years, they have been applied to unfold some neutron energy spectrum. In summary, these methods are all able to unfold neutron energy spectrum. However, the accuracy or efficiency is not as good as expected. Therefore, finding a stable method to unfold neutron energy spectrum is urgent required.

The purpose of this paper is to present a method to unfold neutron energy spectrum based on the potential reduction interior point algorithm. In addition, other types of neutron energy test spectrum are also employed. Solution results showed the good behavior of this algorithm. It is shown to be sufficiently fast, accurate, and robust for measurement of neutron energy spectrum.

## 2. Principle of neutron energy spectrum unfolding

The recoil proton response P(E') of a detector to a neutron energy spectrum N(E) can be written as the first kind Fredholm integral function:

$$P(E') = \int_{E_{max}}^{E_{min}} R(E, E') N(E) dE \qquad (1)$$

where P(E') denotes the recoil proton energy spectrum, N(E) denotes the neutron energy spectrum, R(E,E') is the detector response. The basic task of neutron energy spectrum unfolding is to determine N(E) from the measured recoil proton energy spectrum P(E`) according to the detector response R(E,E`). This is a typical inverse problem which R(E,E`) and P(E`) are known and N(E) is unknown.

If break E and E` into discrete intervals, then the integral equation (1) is rewritten to discrete matrix as follows:

$$P_i = \sum_j R_{ij} N_j \qquad (2)$$

where $N_j$ represents neutron energy spectrum in discrete form, j is the energy group number. $P_i$ is similar. It represents recoil proton energy spectrum in discrete form, i is the energy group number. When the energy group number of i, j is large enough, the equation (2) can replace the equation (1).

In general, the neutron energy spectrum unfolding is very difficult. The response of detector $R_{i,j}$ is usually ill conditioned, which minute perturbation can cause large fluctuations in the solution result. In addition, the first kind Fredholm integral function is a typical ill-posed problem, it has the characteristic of solution result does not exist or exist but is not unique. Therefore, how to accurately solve the ill conditioned in equation (2) is the key problem in the study of neutron energy spectrum unfolding.

## 3. Potential reduction interior point algorithm

Interior point algorithm was first proposed to solve linear programming problems by Karmarkar in 1984 [15]. At present, it has been a very active direction in the field of linear programming. Potential reduction interior point algorithm is developed from interior point algorithm. In this paper, we first apply this algorithm to unfold the neutron energy spectrum.

This algorithm is derived from the reference [11]. Firstly, we transform inverse problem in equation (2) into a least squares problem. Consider a typical least squares:

$$\min_{x \in R} f(x) = \|Ax - b\|^2 \tag{3}$$

here, $A \in R^{m \times n} (m \geq n), b \in R^m, x \in R^n$.

The necessary and sufficient conditions to achieve the optimal solution can be summarized as follows:

$$\begin{cases} x^T(A^TAx - A^Tb) = 0 \\ A^TAx - A^Tb \geq 0 \\ x \geq 0 \end{cases} \tag{4}$$

assume that $M = A^TA, q = -A^Tb$, M is a positive semidefinite matrix, then above conditions can be transformed into a monotone linear complementarity problems(LCP(M,q)):

$$\begin{cases} x^Ty = 0 \\ y = Mx + q \\ x \geq 0, y \geq 0 \end{cases} \tag{5}$$

then, the potential function is constructed:

$$\emptyset(x,y) = (n + \rho) \log(x^Ty) - \sum_{i=1}^n \log(x_iy_i) - n\log(n) \tag{6}$$

$\rho$ is iterative parameter, n is the matrix order.

Note $S_{++} = \{(x, y) : y = Mx + q, x > 0, y > 0\}$ is the strictly feasible interior point for LCP(M,q).

The basic concept of potential reduction interior point algorithm is to adjust the iteration step that satisfy the solution of the linear complementarity problem, at the same time, the potential function value is decreased.

$$\begin{cases} (x + \theta\Delta x, y + \theta\Delta y) \in S_{++} \\ \emptyset(x + \theta\Delta x, y + \theta\Delta y) - \emptyset(x, y) \le -\delta \end{cases} \quad (7)$$

$\delta$ is the descent parameter.

Iteration direction($\triangle$x, $\triangle$y) depends on the solution of Newton direction and central path direction simultaneous equation group:

$$\begin{cases} y\Delta x + x\Delta y = h \\ -M\Delta x + \Delta y = 0 \end{cases} \quad (8)$$

h $= -\left[xy - \beta\frac{x^T y}{n}e\right]$, $\beta = \frac{n}{n+\rho}$. The new iteration direction can reach satisfactory solution with high convergence speed. Termination condition is $(x^k)^T y \le \varepsilon$. At last, the optimal solution is x$^k$.

According to theoretical derivation, when the parameter of the potential function $\rho = O(\sqrt{n})$, the solution of (x$^k$, y$^k$) can be obtained after at most $O\left(\sqrt{n}log\left(\varepsilon^{-1}2^{\emptyset(x_0, \ y_0)/\rho}\right)\right)$ times iteration.

## 4. Simulation and test

In the response matrix, each row corresponds to a given neutron energy and each column corresponds to a given pulse height. To unfold measured neutron energy spectrum, it is required to precisely know the response function R(E,E`) of the detector for each particle energy E. The response matrix can be obtained by Monte Carlo simulation.

Geant4 (for GEometry ANd Tracking) is a package developed by CERN for simulating performance of detectors in nuclear and high energy physics [12]. The response matrix R(E,E`) of the mono-energetic neutrons at the polyethylene film should be confirmed according to the above iteration algorithm.

In order to test and verify this algorithm, the response of system geometry and its physical process are constructed in the Geant4 toolkit. Polyethylene target radius is 5mm, thickness is 0.2mm, neutron source distance 15mm, energy from 1MeV to 16MeV interval 0.1MeV. On the other side of polyethylene target, recoil proton are received in 0° and 30°. The global geometric structure of whole detection system is shown in Figure 1. And then, recoil proton energy spectrum is discrete processing from 0.1MeV to 16 MeV interval 0.1 MeV. Thus the recoil proton energy spectrum response matrix of 0° and 30° are obtained.

The program of potential reduction interior point algorithm is prepared by MATLAB. The response matrix and recoil proton energy spectrum are brought into the program, and then we get the solution results. Several neutron energy spectrums with mono-energetic, multi-peak, gauss and double gauss as the output data to test the unfolding capability.

## 5. Results and discussion

Figure 2~5 show the solution results of the unfolded energy spectrum. Figure 2 shows the solution results of gauss neutron energy spectrum with $\sigma$ =0.2 MeV, $\mu$ =11.0MeV. Figure 3 shows the solution results of neutron energy spectrum with 4 peaks at 5, 7, 9 and 11 MeV of which intensity ratio is 1:2:3:4. Figure 4 shows the solution results of double gauss neutron energy spectrum with $\sigma_1$=0.15 MeV, $\mu_1$=9.0 MeV, $\sigma_2$=0.15 MeV, $\mu_2$=13.0 MeV. Figure 5 shows the solution results of neutron energy spectrum with 1 peak at 9.81 MeV.

We defined Q to evaluate the unfolded solution results [13]. Obviously, a perfect unfolded result would match exactly with the true spectrum and gives Q=0. The Q values are list in Table.1.

$$Q = \left[ \frac{\sum_{i=1}^{n}(N_{i,Solution} - N_{i,True})^2}{\sum_{i=1}^{n}(N_{i,True})^2} \right]^{1/2} \tag{9}$$

As can be seen from Figure 2~5 and Table 1, whatever type of energy spectrum, the approximate results can be obtained in each direction, and the results of 0 ° are more prominent. Its Q value is smaller than 30 ° results. The solution results of 30 °about 10% fluctuations compared with the true. The solution results in low energy region has a little inaccurate. It may be that the response matrix is not accurate enough in this region. Overall, the results of 0 ° are reliable, it can be applied to the solution of neutron energy spectrum directly. The solution results of 30 ° is not as accurate as the solution results of 0 °.

The prominent advantage is that it has been accelerating the convergence speed of reach to the optimal solution. It has a relatively high efficiency, compared with the general algorithm such as GRAVEL. Take the solution results of gauss neutron energy spectrum in Figure 2 as an example. To achieve the same accuracy while all other conditions are identical., the GRAVEL algorithm takes 26794 steps while the potential reduction interior point algorithm takes only 13238 steps. The unfolding efficiency is increased by nearly half.

The disadvantage of the potential reduction interior point algorithm is that it is sensitive to the response matrix. As shown in Figure 4 and Figure 5, in the solution results of 30 °, some outliers from the true solution are observed at low energy range (from 2 to 4 MeV). It is not obvious in the solution results of 0 °. This is due to the outgoing direction of the recoil proton is concentrated in the forward direction, the response matrix of 0 ° is more accurate than the 30 ° direction.

In conclusion, the potential reduction interior point algorithm can obtain satisfactory solution results with faster convergence speed. When the case of response matrix is accurate, this unfolding algorithm is stable and provide efficiency accurate results.

As an application example, Figure 6 shows the solution result of D-T fusion neutron energy spectrum. This work can provide support for fusion neutron energy spectrum diagnosis. The data are from reference [13]. The full width at half maximum (FWHM) of the solution result and the true D-T fusion neutron energy spectrum is 1.1225MeV and 1.1957 MeV respectively.

## 6. Conclusion

In this paper, an algorithm based on the potential reduction interior point has been developed and applied to the neutron spectrum unfolding. This algorithm selects the appropriate step length in each iteration process so that the value of potential function decreased, and the iteration direction is the simultaneous solution of central path direction and newton direction. So, it can be concluded that this algorithm is sufficiently fast and the solution results match up with the expected results. Additionally, solution results of the detector lied at 0 ° are better than the detector lied at 30 °. Further work will focus on how to get more accurate results by using the data of multiple direction, and a new method called multi-directional detection scheme will be discussed.

## Acknowledgements

Financial supports from the Fundamental Research Funds for the Central Universities (Grant No.2016XS61), and the Opening Foundation of State Key Laboratory of Intense Pulsed Radiation Simulation and Effect are gratefully acknowledged.

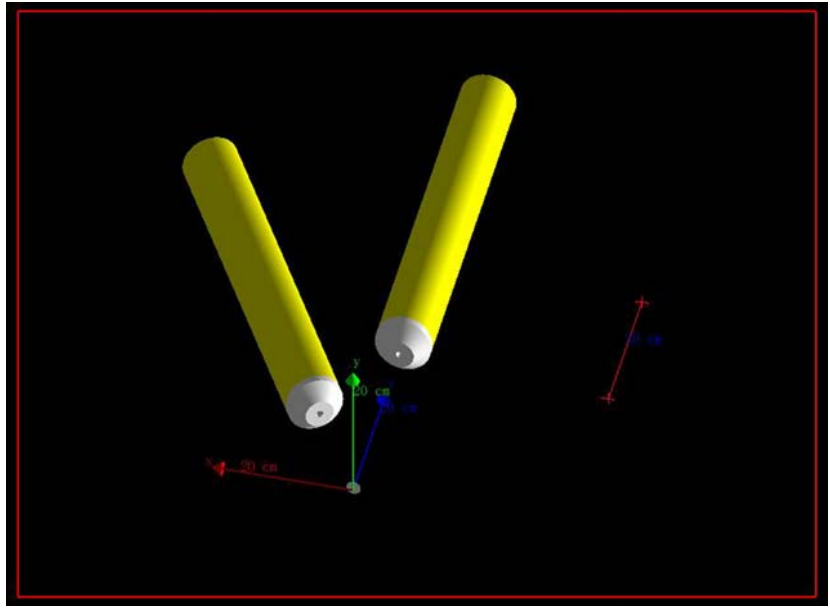

Fig.1 Global geometric structure of whole detection system. The recoil proton detectors are placed in 0 ° and 30 °. Thickness of Polyethylene film is 18.4 mg/cm$^2$.

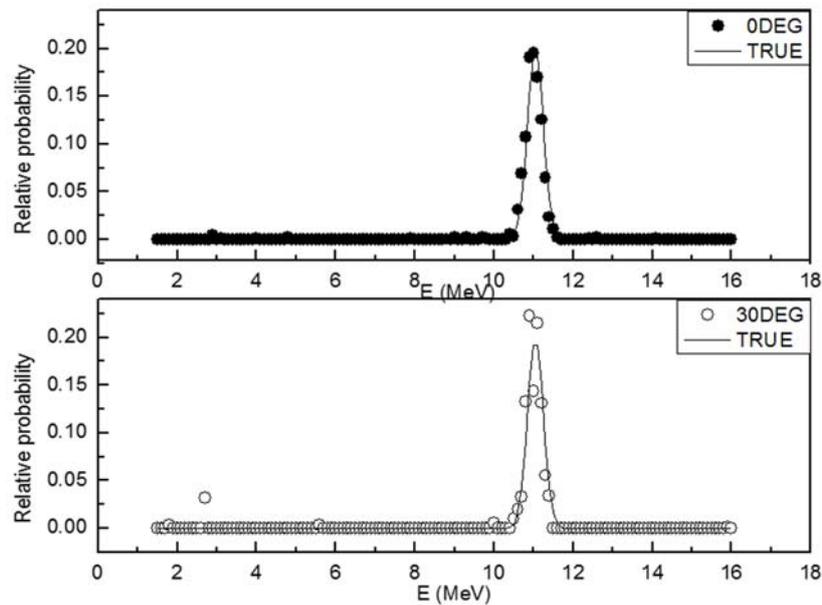

Fig.2 Solution results of gauss neutron energy spectrum with $\sigma$ =0.2 MeV, $\mu$ =11.0MeV. Solid points represent the 0 ° results, circled points represent the 30 ° results and solid line is the true neutron energy spectrum.

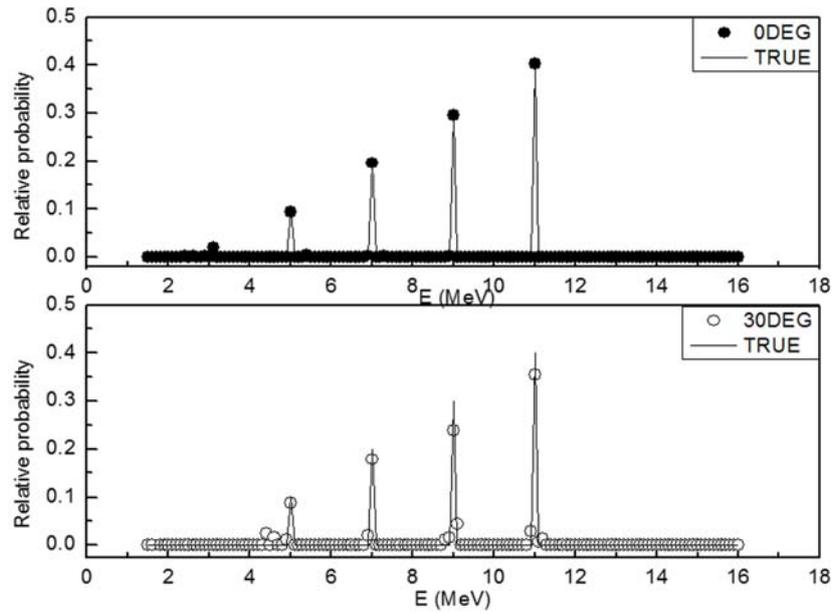

Fig.3 Solution results of neutron energy spectrum with 4 peaks at 5, 7, 9 and 11 MeV of which intensity ratio is 1:2:3:4. Solid points represent the 0 ° results, circled points represent the 30 ° results and solid line is the true neutron energy spectrum.

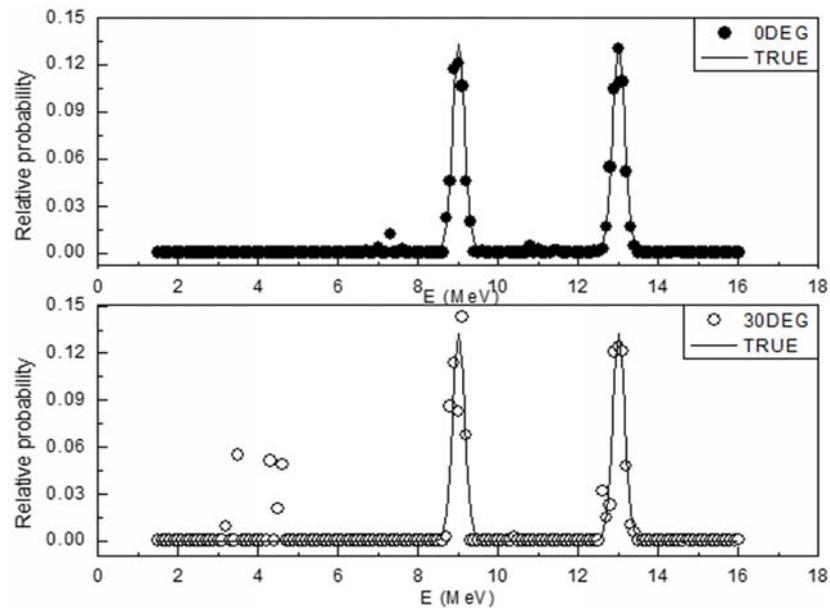

Fig.4 Solution results of double gauss neutron energy spectrum with $\sigma_1$=0.15 MeV, $\mu_1$=9.0 MeV, $\sigma_2$=0.15 MeV, $\mu_2$=13.0 MeV. Solid points represent the 0 ° results,

circled points represent the 30 ° results and solid line is the true neutron energy spectrum.

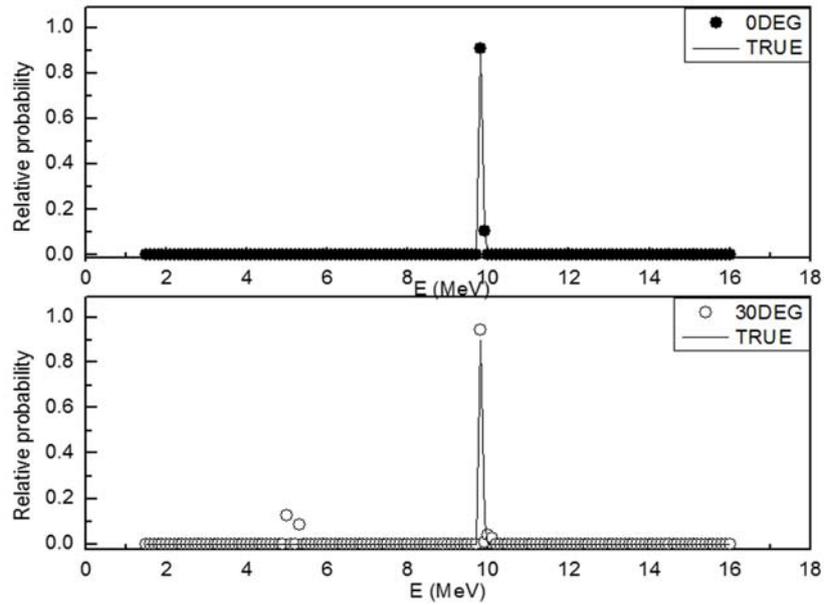

Fig.5 Solution results of neutron energy spectrum with 1 peak at 9.81 MeV. Solid points represent the 0 ° results, circled points represent the 30 ° results and solid line is the true neutron energy spectrum.

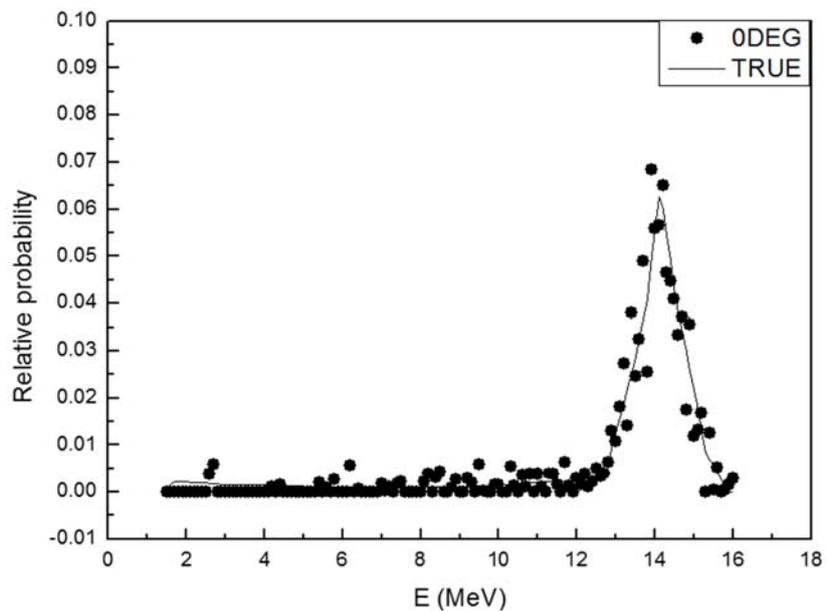

Fig.6 Solution results of D-T fusion neutron energy spectrum corresponding to the detector at 0°. The true data from reference [13]. Solid points represent the 0 ° results, and solid line is the true D-T fusion neutron energy spectrum.

Table 1 Evaluation of unfolded results for spectrums in Figure 2 to 5. Q is defined in equation (9). The closer that this value is to 0, the more excellent unfolded result is.

|  | Figure 2 | Figure 3 | Figure 4 | Figure 5 |
|---|---|---|---|---|
| $Q(0°)$ | 0.2304 | 0.0562 | 0.1041 | 0.0235 |
| $Q(30°)$ | 0.3393 | 1.0594 | 0.6338 | 0.4097 |